\documentclass[11pt,a4paper]{article}
\usepackage{macrofile}
\usepackage{formatfile}
\usepackage{natbib}
\usepackage[nomarkers,nolists]{endfloat}
\newcommand{\epV}{\mathbf{V}}
\newcommand{\epU}{\mathbf{U}}
\newcommand{\bX}{\mathbf{X}}

\newcommand{\Kac}{\mathscr{K}}
\newcommand{\Gauss}{\mathscr{N}}
\newcommand{\Dcong}{\stackrel{\mathcal{D}}{\simeq}} 

\title{Uniform reliability tests for forecasting systems with small lead time}
\author{J.~Br\"{o}cker}
\begin{document}
\maketitle
\begin{abstract}	
  A long noted difficulty when assessing the reliability (or calibration) of forecasting systems is that reliability, in general, is a hypothesis not about a finite dimensional parameter but about an entire functional relationship.
A calibrated probability forecast for binary events for instance should equal the conditional probability of the event given the forecast for {\em any} value of the forecast.
Attempts to estimate deviations from calibration at a specific forecast value meet with the difficulty that the probability of the forecast assuming that value is typically zero.
Considering the estimated {\em cumulative} deviations from reliability instead however, tests are presented for which the asymptotic distribution of the test statistic can be established rigorously.
The distribution turns out to be universal, provided the forecasts ``look one step ahead'' only, or in other words, verify at the next time step in the future.
Furthermore, the tests develop power against a wide class of alternatives.
Numerical experiments for both artificial data as well as operational weather forecasting systems are also presented, as are possible extensions to forecasts with longer lead times.
\end{abstract}
%
%
%
%
\section{Introduction}
\label{sec:introduction}
A probability forecast (for binary or ``yes~vs~no'' events) is called reliable if the probability of the event, conditionally on the forecast taking a specific value, is equal to that same value;
this should hold for {\em any} value the forecast may assume.
Similar definitions of reliability referring to conditional mean or conditional quantile forecasts exist.
In the context of an operational forecasting system for real world variables (e.g.\ environmental or economic), reliability can only be assessed in a statistical sense.
That is, provided with an archive of verification--forecast pairs, we may formulate reliability as a statistical hypothesis and perform statistical tests for that hypothesis.
Typically though, reliability is a hypothesis not about a finite dimensional parameter but about an entire functional relationship, a long noted difficulty when assessing the reliability of forecasting systems.
Tests for that hypothesis based on estimated deviations from that functional relationship for specific values of the forecast meet with the difficulty that the probability of the forecast assuming any specific value may be zero.
A possible remedy is to consider weaker forms of reliability instead, for instance that the average of the forecast agrees with the average of the verification.
This is but a necessary consequence of full reliability, and forecasting systems exhibiting this weak or unconditional form of reliability are generally inadequate for decision support.
The aim of the present paper is to discuss tests that use an estimate of cumulative deviations from reliability and are thereby able to take the functional character of the reliability hypothesis fully into account.
Based on a rigorous methodology, the asymptotic distribution of the test statistic is established and turns out to be universal, that is, independent of the specifics of the underlying data source.
The fact that the verification--forecast pairs are not temporarily independent is taken into account, the only structural assumption being that forecasts always verify at the next time step in the future.
Relevant concepts and notation will be introduced and made precise in Section~\ref{sec:reliability}.
In particular, the concept of reliability for probability forecasts (of binary events), for conditional mean forecasts, and for conditional quantile forecasts, will be discussed.
Section~\ref{sec:methodology} discusses the tests, with detailed instructions as to how to perform the tests as well as the asymptotic properties.
The material in that section is based on a rigorous mathematical analysis we deferred to Appendix~\ref{apx:uclt} for the interested reader.
Section~\ref{sec:numerical_experiments} contains a number of numerical experiments, both for artificial data as well as forecasts from an operational weather forecasting centre.
The main purpose of these experiments is to demonstrate the feasibility of the methodology, and to discuss and illustrate further practical aspects such as the amount of required data.
The question of test power is discussed in Section~\ref{sec:power}, where a mathematical analysis demonstrates the ability of the tests to develop power against a wide class of alternatives, supplemented by numerical results.
Section~\ref{sec:conclusions} concludes and discusses further avenues of research, in particular regarding other types of forecasts as well as forecasts with larger lead times.
%
%
\section{Reliability of probability, mean, and quantile forecasts}
\label{sec:reliability}
To discuss the problem of reliability more formally, let $\{Y_k, k = 1, 2, \ldots \}$ be a series of verifications (i.e.\ observations), which in the present paper we assume to be random variables having values either in the real numbers or just in the binary set $\{0, 1\}$.
These two cases will be referred to as the continuous and the binary case, respectively.
We consider corresponding forecasts $\{f_k, k = 1, 2, \ldots\}$ which are real numbers in both cases.
Our tests will be based on the joint data $\{(Y_k, f_k), k = 1, 2, \ldots\}$ to which we refer as the {\em forecast--verification pairs}.
The index $k$ is a temporal index or time step, and the forecast $f_k$ corresponds to the verification $Y_k$ which obtains at some point $t_k$ in actual time.
Typically, the forecast $f_k$ is issued at some point prior to $t_k$, and the lag $L_k$ between the time when $f_k$ is issued and the time it verifies (i.e.\ $t_k$) is often referred to as the {\em lead time}.
Although the lead time is often independent of $k$, there are examples where this is not the case, for instance in seasonal forecasting systems that focus on specific periods of the year; we will however assume the lead time to be constant and drop the subscript on $L$ for notational simplification.
A fundamental assumption of this paper is that the lead time $L$ is never larger than $t_k - t_{k-1}$, or if measured in time steps rather than absolute time, the lead time is equal to (or smaller than) one.
(This is the ``small lead time'' condition alluded to in the title of this paper.)
We can therefore assume that when issuing the forecast $f_k$, the forecaster has access not only to all previous forecasts but also to all previous verifications, that is, she knows $(Y_1, \ldots, Y_{k-1})$.
Therefore, this information can in principle factor into the forecast.
The concept of {\em reliability} (or {\em calibration}) refers to how the forecasts are actually to be interpreted.
We start with the binary case.
Here, a common interpretation of reliability is that for each $k$ the forecast $f_k$ is equal to the probability of the event $Y_k = 1$, conditional on the forecast $f_k$ itself.  
Mathematically, we might express this as
\beq{equ:1.10}
\P(Y_k = 1 | f_k) = f_k
\qquad \text{for all $k = 1, 2, \ldots$}.
\eeq
This definition of reliability though makes no reference to (temporally) previous forecasts and verifications.
To develop tests however and rigorously establish their statistical properties, the temporal dependencies of the verification-forecast pairs have to be taken into account.
Rather than introducing specific assumptions concerning these dependencies ad hoc (or worse still, tacitly), the approach of the present paper is to obtain all necessary information regarding these temporal dependencies from the reliability assumption itself, potentially at the expense of using a stronger but nonetheless well motivated reliability hypothesis.
As noted previously, the forecaster has access to all previous forecasts and verifications when issuing the forecast $f_k$, and in an ideal world, we may assume that the forecaster has taken this information into account, and that the current forecast represents the best possible one given this information.
Therefore, we would expect that for each $k$ the forecast $f_k$ is equal to the probability of the event $Y_k = 1$, conditional on the current forecasts as well as {\em all previous} forecasts and observations.
We might write this as
\beq{equ:1.20}
\P(Y_k = 1 | f_{1:k}, Y_{1:k-1}) = f_k
\qquad \text{for all $k = 1, 2, \ldots$}
\eeq
where here (and in the following) we use the shorthand $Y_{k:l} := (Y_k, \ldots, Y_l)$ for any $k \leq l$.
Condition~\eqref{equ:1.20} implies condition~\eqref{equ:1.10} as can be shown using elementary probability calculus.
Condition~\eqref{equ:1.20} will constitute our null hypothesis, and the main aim of the present paper is to develop statistical tests for this hypothesis.
We need to emphasise however that even though the null hypothesis~\eqref{equ:1.20} is assumed throughout our analysis for binary forecasts, the tests are expected to develop power only against alternatives to the restricted hypothesis~\eqref{equ:1.10}.
This is due to the great generality of the former hypothesis.
We stress that to the best of our knowledge, the presented tests are the first that can be rigorously shown to develop power against a wide range of alternatives to hypothesis~\eqref{equ:1.10}.
There are several different ways to generalise the reliability hypothesis~\eqref{equ:1.20} to the continuous case, depending on how the forecasts $\{f_k\}_{k \in \N}$ are to be interpreted.
We will focus on mean forecasts and quantile forecasts.
In the first case, we basically just replace conditional probabilities with conditional expectations, that is, we require
\beq{equ:1.30}
\E(Y_k| f_{1:k}, Y_{1:k-1}) = f_k
\qquad \text{for all $k = 1, 2, \ldots$},
\eeq
or equivalently that the conditional expectation of $Y_k$ given past verifications as well as current and past forecasts is given by the current forecast $f_k$ for all $k = 1, 2, \ldots$.
In the case of quantiles of a fixed level $\alpha$, say, we require that
\beq{equ:1.40}
\P(Y_k \leq f_k| f_{1:k}, Y_{1:k-1}) = \alpha
\qquad \text{for all $k = 1, 2, \ldots$}.
\eeq
Regarding power, a remark similar to the one made for the binary case applies to both conditional mean and conditional quantile forecasts.
Even though hypothesis~\eqref{equ:1.30} resp.~\eqref{equ:1.40} are imposed in the conditional mean resp.\ conditional quantile forecasts, power will only be demonstrated against equivalent forms of the restricted hypothesis~\eqref{equ:1.10}, namely against
\beq{equ:1.43}
\E(Y_k | f_k) = f_k
\qquad \text{for all $k = 1, 2, \ldots$}
\eeq
in the conditional mean case, and
\beq{equ:1.47}
\P(Y_k \leq f_k | f_k) = \alpha
\qquad \text{for all $k = 1, 2, \ldots$}
\eeq
in the quantile case.
We will finish this section with a discussion as to how these conditions for reliability need to be modified for larger lead times, and why in that case a rigorous testing methodology is harder to develop.
As discussed above, the lead time is, roughly speaking, the lag between the time when the forecast is issued and when the verification obtains.
This means that if the forecast $f_k$ has a lead time $L$, say, then at the time the forecast is issued, only the verifications $Y_1, \ldots, Y_{k - L}$ are available to the forecaster while the verifications $Y_{k - L + 1}, \ldots, Y_k$ are still in the future.
Therefore in the conditioning in hypotheses~(\ref{equ:1.20},\ref{equ:1.30},\ref{equ:1.40}), we merely have to replace the verifications $Y_{1:k}$ with $Y_{1:k-L}$.
Hypothesis~\eqref{equ:1.10} for instance will now read as 
\beq{equ:1.50}
\P(Y_k = 1| f_{1:k}, Y_{1:k-L}) = f_k
\qquad \text{for all $k = 1, 2, \ldots$}.
\eeq
As discussed, a difficulty in developing tests for reliability in general lies in the fact that the verification--forecast pairs are dependent random variables, with the only a~priori information about the nature of the dependencies being the reliability hypothesis itself.
In that regard, the hypothesis~\eqref{equ:1.20} for the case of lead time $L=1$ provides a lot more information than the corresponding hypothesis~\eqref{equ:1.50} for larger lead times.
As a consequence, the statistical properties of the test can be derived in the case of lead time $L=1$ under minimal additional assumptions, while we expect that further assumptions will be required for the case of higher lead times.
This is entirely analogous to the difficulties one faces when testing reliability of ensemble forecasting systems~\citep[see][]{broecker_stratified_serial_dependence_2020}.
Having said this, the core result in Appendix~\ref{apx:uclt} holds true even in the case of larger lead times if a mixing condition is imposed, and the remaining difficulties do not seem insurmountable (see the conclusions for further discussion).
%
%
\section{Methodology and main results}
\label{sec:methodology}
In this section, we will motivate and discuss the test statistics and present the main results regarding the properties of the tests.
These results can be made entirely rigorous, although the proofs require considerable mathematical machinery.
Some of the details are discussed in Appendix~\ref{apx:uclt} for the interested reader.
From now on, we make the following assumptions (these will be made precise in Appendix~\ref{apx:uclt} and augmented by several inexorability conditions):
\begin{enumerate}
\item the reliability hypothesis~\eqref{equ:1.20} (or~\eqref{equ:1.30} resp.~\eqref{equ:1.40} for the conditional mean resp.\ conditional quantile case) is in force;
\item the verification--forecast pairs $\{(Y_k, f_k), k = 1, 2, \ldots\}$ form a stationary and ergodic process;
\item the distribution of the forecast $f_k$ conditionally on $(Y_l, f_l)$ for $l = 1, \ldots, k-2$ is continuous (to be made precise in Appendix~\ref{apx:uclt}).
\end{enumerate}
By stationarity we mean that for any $l$, the joint distribution of
\[
\big( (Y_{k+1}, f_{k+1}), \ldots, (Y_{k+l}, f_{k+l}) \big)
\]
does not depend on $k$.
As a consequence, everything that depends on the distributions such as moments, pair correlations etc is independent of time.
Ergodicity means (for the purpose of this paper), that time averages converge to expected values. 
%
%
\subsection{Forecasts for binary verifications}
\label{sec:binary_forecasts}
One of the most popular tools to assess the reliability of binary forecasts is the reliability diagram.
If we write hypothesis~\eqref{equ:1.10} as
\beq{equ:2.10}
\P(Y_k = 1 | f_k = p) = p
\qquad \text{for all $k = 1, 2, \ldots; p \in [0, 1]$},
\eeq
and estimating the left hand side for several values of $p$ and plotting those estimates versus $p$~gives a reliability diagram~\citep[see for instance][]{wilks95,atger04-1,broecker06-4,broecker_chapter_2011}.
It should exhibit a graph close to the diagonal, up to ``random fluctuations'', provided the forecasting system is reliable.
This however requires dividing the range of $p$ up into several bins in a somewhat arbitrary fashion.
We aim to remove this requirement and construct a test that assesses reliability {\em uniformly} across all values of the forecast, that is, the entire unit interval.
To motivate our test statistic, we integrate Equation~\eqref{equ:2.10} over $p \in [0, \zeta]$ against the distribution function $F$ of $f_k$.
(Note that by stationarity, $F$ does not depend on $k$.)
We obtain
\beq{equ:2.20}
\P(Y_k = 1, p \leq \zeta) = \int_0^{\zeta} p \idd F(p)
\eeq
The hypotheses~\eqref{equ:2.10} and~\eqref{equ:2.20} are entirely equivalent.
If we assume that a law of large numbers holds, we could estimate the difference between both sides of hypothesis~\eqref{equ:2.20} by empirical averages, which gives
\beqn{equ:2.25}
\epU_n(\zeta) := \frac{1}{n} \sum_{k = 1}^n (Y_k - f_k) \* \cf_{\{f_k \leq \zeta\}}.
\eeq
(By $\cf_{\{A\}}$ we denote the indicator function of the event $A$).
We would then expect that for every $\zeta \in [0, 1]$ the random quantity $\epU_n(\zeta)$ should be small for large $n$.
If in addition a central limit theorem holds, we might even say how small, namely $\sqrt{n} \epU_n(\zeta)$ would be normally distributed with mean zero and a certain variance.
The variance of this quantity can be calculated from its definition using hypothesis~\eqref{equ:1.20}, but essentially the same calculations give the following more general result:
Defining the function $G(\zeta) := \int_0^{\zeta} p(1-p) \idd F(p)$ we find
\beq{equ:2.30}
\E \big( n \epU_n(\xi) \epU_n(\eta) \big)
 = G(\xi \wedge \eta),
\eeq
where $a \wedge b$ denotes the minimum of $a$ and $b$.
For each $n \in \N$, we might thus regard $\sqrt{n} \epU_n$ as a stochastic process in the continuous parameter $\zeta \in [0, 1]$; this process has mean zero and covariance function given by Equation~\eqref{equ:2.30}.
Note that the covariance function is independent of $n$.
Our test will be based on the highly nontrivial fact that the law of large numbers and the central limit theorem hold {\em uniformly} in the parameter $\zeta$.
(The precise result is stated in Appendix~\ref{apx:uclt}.)
The limit in distribution of $\sqrt{n} \epU_n$ as $n \to \infty$ is given by a Gaussian process on the unit interval with mean zero and covariance function given by Equation~\eqref{equ:2.30}.
This limiting process, which we will denote with $\{ \epU(\zeta), \zeta \in [0, 1] \}$, can be described by means of the Wiener process (aka standard Brownian motion).
The Wiener process $\{W(t); t \in [0, 1]\}$ is a Gaussian process with mean zero and covariance function given by $\E W(t) W(s) = t \wedge s$.
Therefore, we find that $\{W(G(\zeta)), \zeta \in [0, 1]\}$ is a Gaussian process with mean zero and covariance function
\beqn{equ:2.40}
\E W(G(\xi)) W(G(\eta))
= G(\xi) \wedge G(\eta)
= G(\xi \wedge \eta),
\eeq
the last equality being true because $G$ is monotonically increasing.
Comparing with Equation~\eqref{equ:2.30} we find $\epU(\zeta) = W(G(\zeta))$ for all $\zeta \in [0, 1]$ (the equality sign here means that both processes have the same distribution).
As a (preliminary) candidate for a test statistic, we consider
\beq{equ:2.50}
\tau_n := \sup_{\zeta \in [0, 1]} |\epV_n(\zeta)|
\qquad \text{with} \qquad
\epV_n(\zeta) := \sqrt{\frac{n}{G(1)}} |\epU_n(\zeta)|.
\eeq
Due to the central limit theorem being uniform, we can conclude that for $n$ large
\beqn{equ:2.60}
\tau_n
\Dcong \sup_{\zeta \in [0, 1]} \frac{1}{\sqrt{G(1)}} |W(G(\zeta))|
= \sup_{\zeta \in [0, 1]} |W(\zeta)|.
\eeq
Here $\Dcong$ means that the distributions are approximately the same.
The distribution of the supremum of the Wiener process is well known, see~\citet{erdos_certain_limit_theorems_1946}; in the following, the symbol~$\Kac$ will denote the cumulative distribution function.
Strictly speaking, $\tau_n$ as in Equation~\eqref{equ:2.50} is not a test statistic as it contains the unknown factor $G(1) = \E f_k (1 - f_k)$.
This factor though can be replaced with an estimator, for instance an empirical average.
In summary, the suggested test comprises the following steps:
(A python package containing code for this test---as well as for mean and quantile forecasts discussed below---is available online~\citep{franz_2020};
this code has been used throughout this paper.)
\begin{enumerate}
  \item compute $\epU_n$ according to Equation~\eqref{equ:2.30};
  \item estimate $G(1)$ through $\gamma_n := \frac{1}{N} \sum_{k = 1}^n f_k (1 - f_k)$;
  \item find $\tau_n = \sup_{\zeta \in [0, 1]} \sqrt{\frac{n}{\gamma_n}} |\epU_n(\zeta)|$ (this step essentially requires sorting over the values $f_1, \ldots, f_n$ rather than optimisation as the supremum will be assumed at one of these values);
  \item finally, reject the null hypothesis if $\pi := 1 - \Kac(\tau_n) $ exceeds the chosen significance level, otherwise accept (or report $\pi$ as the $p$--value).
\end{enumerate}
%
%
%
As an aside, one might wonder how the proposed test can possibly work given the difficulties identified in~\citet{atger04-1,broecker07-1} with ``continuous parameter'' reliability diagrams.
After all, the test assesses the deviations from the perfectly diagonal reliability diagram uniformly for all values of the forecast.
It is however the {\em cumulative} deviations that form the test statistic.
Under the null hypothesis, these behave basically like a Wiener process.
The deviations themselves would consequently be (roughly speaking) the derivative of the Wiener process which behaves like white noise.
Working with white noise directly though is mathematically much harder, one of the problems being that white noise realisations are not functions.
By considering the cumulative deviations instead we avoid these difficulties.
\subsection{Conditional mean forecasts}
\label{sec:mean_forecasts}
In the case of mean forecasts, entirely analogous considerations will lead us to the test statistic.
We now use hypothesis~\eqref{equ:1.43} instead of~\eqref{equ:1.10} and write it as
\beq{equ:2.65}
\E(Y_k | f_k = z) = z
\qquad \text{for all $k = 1, 2, \ldots; z \in \R$.}
\eeq
Similar to the binary case, we integrate Equation~\eqref{equ:2.65} over $z \in [-\infty, \zeta]$ against the distribution function $F$ of $f_k$. (Note that by stationarity, $F$ does not depend on $k$.) We obtain
\beq{equ:2.70}
\E(Y_k | f_k \leq \zeta) = \int_{-\infty}^{\zeta} z \idd F(z).
\eeq
The hypotheses~\eqref{equ:2.65} and~\eqref{equ:2.70} are entirely equivalent, and by arguments entirely analogous to the binary case we are lead to consider
\beq{equ:2.80}
\epU_n(\zeta) := \frac{1}{n} \sum_{k = 1}^n (Y_k - f_k) \* \cf_{\{f_k \leq \zeta\}},
\eeq
which is the same expression as $\epU_n$ in Equation~\eqref{equ:2.20} except that now $\zeta$ ranges over the whole of $\R$ instead of only the unit interval.
Defining the function $G(\zeta) := \int_0^{\zeta} \sigma(z) \idd F(z)$, where $\sigma(z) = \E((Y_k - z)^2|f_k = z)$ is the conditional variance, we find
\beq{equ:2.90}
\E \big( n \epU_n(\xi) \epU_n(\eta) \big)
 = G(\xi \wedge \eta).
\eeq
Again, our test will be based on a central limit theorem holding uniformly in the parameter $\zeta$.
The limit in distribution of $\sqrt{n} \epU_n$ as $n \to \infty$ is given by a Gaussian process on the unit interval with mean zero and covariance function given by Equation~\eqref{equ:2.90}.
As before, we find that $\{W(G(\zeta)), \zeta \in \R\}$ possesses the relevant distribution.
As a test statistic, we will use
\beqn{equ:2.100}
\tau_n := \sup_{\zeta \in \R} |\epV_n(\zeta)| := \sup_{\zeta \in \R} \sqrt{\frac{n}{\gamma_n}} |\epU_n(\zeta)|, 
\eeq
where $\gamma_n := \frac{1}{n} \sum_{k = 1}^{n} (Y_k - f_k)^2$ serves as an estimator for $G(\infty) = \E((Y_k - f_k)^2)$.
Thanks to the uniform central limit theorem, we can conclude that for $n$ large
\beqn{equ:2.110}
\tau_n  = \sup_{\zeta \in \R} \sqrt{\frac{n}{\gamma_n}} |\epU_n(\zeta)|
\Dcong \sup_{\zeta \in [0, 1]} |W(\zeta)|.
\eeq
In summary, the suggested test comprises exactly the same steps as for the binary case.
%
%
\subsection{Conditional quantile forecasts}
\label{sec:quantile_forecasts}
Our discussion of the quantile forecasts can be very brief.
We write hypothesis~\eqref{equ:1.47} in the equivalent form
\beqn{equ:2.120}
\P(Y_k \leq f_k| f_k = z) = \alpha
\qquad \text{for all $z \in \R$} 
\eeq
and integrate over $z \in [-\infty, \zeta]$ against the distribution function $F$ of $f_k$, we obtain
\beqn{equ:2.140}
\P(Y_k \leq f_k, f_k \leq \zeta) = \alpha \P(f_k \leq \zeta).
\eeq
Estimating both sides through empirical averages, we are lead to consider
\beq{equ:2.150}
\epU_n(\zeta) := \frac{1}{n} \sum_{k = 1}^n (\cf_{\{Y_k \leq f_k\}} - \alpha) \* \cf_{\{f_k \leq \zeta\}}.
\eeq
This time, we find
\beqn{equ:2.160}
\E \big( n \epU_n(\xi) \epU_n(\eta) \big)
 = \alpha(1 - \alpha) \* F(\xi \wedge \eta).
\eeq
Similar to the previous cases, $\{W(\alpha(1 - \alpha) F(\zeta)), \zeta \in \R\}$ provides the limiting distribution of $\sqrt{n} \epU_n$.
As a test statistic, we will use
\beq{equ:2.170}
\tau_n := \sup_{\zeta \in \R} |\epV_n(\zeta)|  := \sup_{\zeta \in \R} \sqrt{\frac{n}{\alpha (1 - \alpha)}} |\epU_n(\zeta)|, 
\eeq
and thanks to the uniform central limit theorem, we can conclude that for $n$ large
\beqn{equ:2.180}
\tau_n 
\Dcong \sup_{\zeta \in [0, 1]} |W(\zeta)|.
\eeq
This time, no additional quantity needs estimating.
%
%
\section{Numerical experiments}
\label{sec:numerical_experiments}
%
%
\newcommand{\nrmcrepeat}{1000}
\newcommand{\nrtstamps}{730}
\newcommand{\pbiasbinaryexp}{0.004}
\newcommand{\pbiasmeanexp}{0.007}
\newcommand{\pbiasquantexp}{0.001}
\newcommand{\pbinaryexp}{0.258}
\newcommand{\pmeanexp}{0.506}
\newcommand{\pquantexp}{0.433}
\newcommand{\pMeanExpLargeNorm}{0.002}
\newcommand{\pMeanExpLargeUnif}{0.468}
In this section, we will present a few numerical experiments for artificial data with the null hypothesis either valid or violated in a controlled fashion, and for data from an operational forecasting system.
All experiments were done using the mentioned python package~\texttt{franz}~\citep{franz_2020}.
\subsection{Artificial data---AR process}
\label{sec:ar_data}
\begin{figure}[p]
  \begin{center}
    (a) \parbox[t][2.1in][c]{3.2in}{\includegraphics{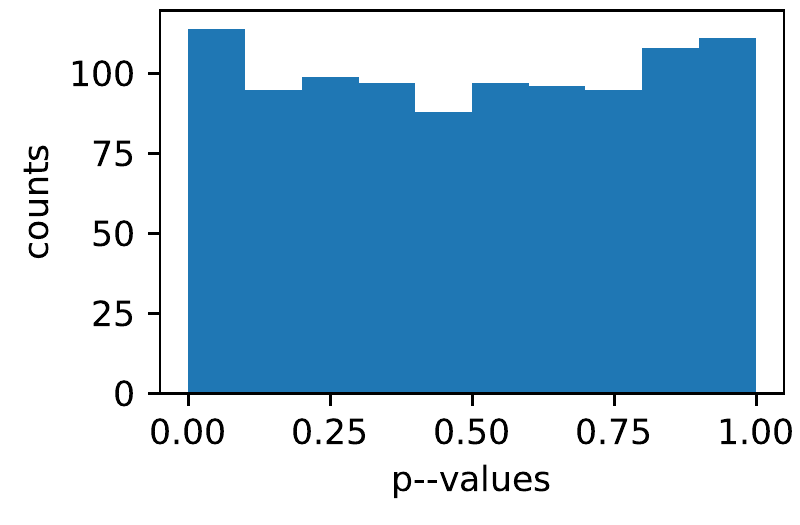}}\\
    (b) \parbox[t][2.1in][c]{3.2in}{\includegraphics{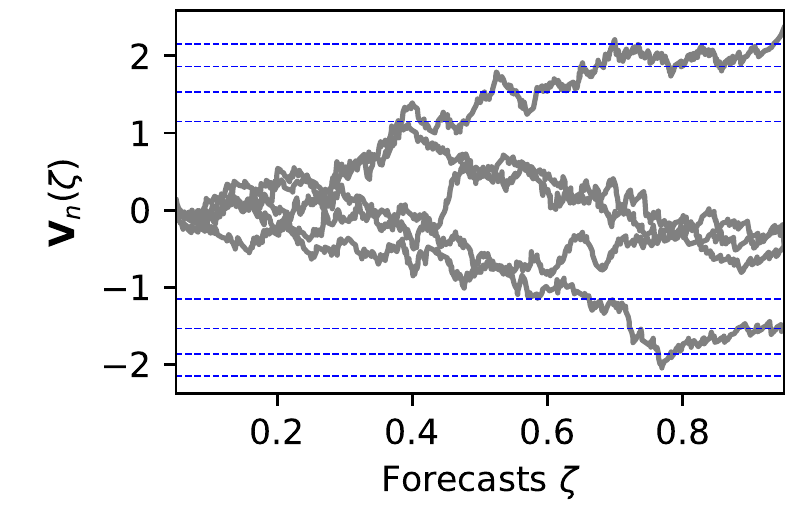}}
  \end{center}
    \caption{\label{fig:binary}
      Test results for binary forecasts for artificial data.
      A Monte-Carlo experiment was carried out, comprising \nrmcrepeat~independent runs.
      In each run, the test was applied to \nrtstamps~verification-forecast pairs.
      Panel (a)~shows a histogram of the $p$-values.
      The histogram shows a uniform distribution, up to statistical fluctuations; this is confirmed through a Kolmogorov-Smirnov test which gives a $p$-value of~\pbinaryexp.
      Panel (b)~shows five typical realisations of the process $\epV_n$.
      The dashed lines in Panel~(b) indicate nested regions that
      the path of $\epV$ leaves with probability $\frac{1}{2}$ (innermost), $\frac{1}{4}, \frac{1}{8}$ and~$\frac{1}{16}$ (outermost).
See text for details.}
\end{figure}

\begin{figure}[p]
  \begin{center}
    (a) \parbox[t][2.1in][c]{3.2in}{%
      \includegraphics{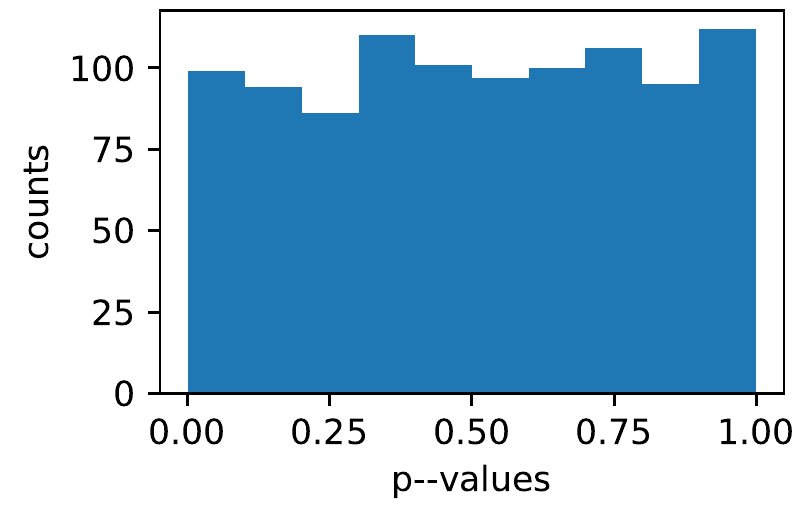}}\\
    (b) \parbox[t][2.1in][c]{3.2in}{%
      \includegraphics{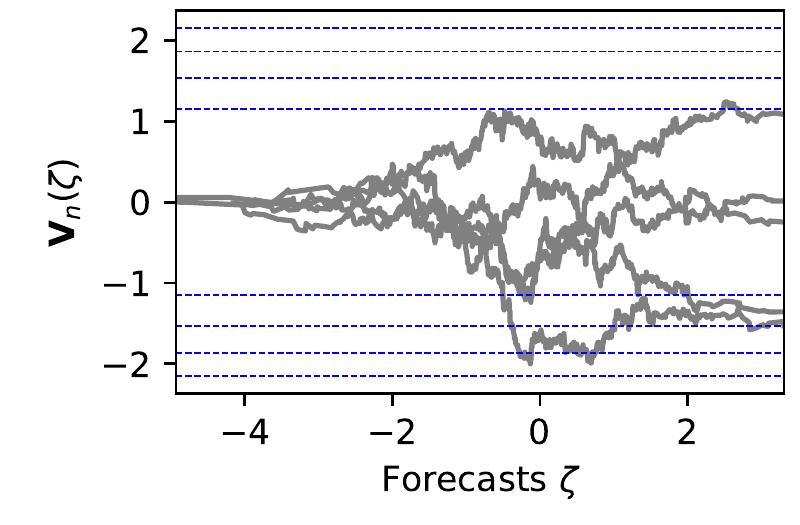}}
  \end{center}
    \caption{\label{fig:mean}
      Test results for mean forecasts for artificial data.
      The setup is exactly as in Figure~\ref{fig:binary}.
      A Kolmogorov-Smirnov test for uniformity of the histogram gives a $p$-value of~\pmeanexp.
      Panel (b)~shows five typical realisations of the process $\epV_n$.
      The dashed lines in Panel~(b) are as in Figure~\ref{fig:binary}.
    }
\end{figure}

\begin{figure}[p]
  \begin{center}
    (a) \parbox[t][2.1in][c]{3.2in}{%
      \includegraphics{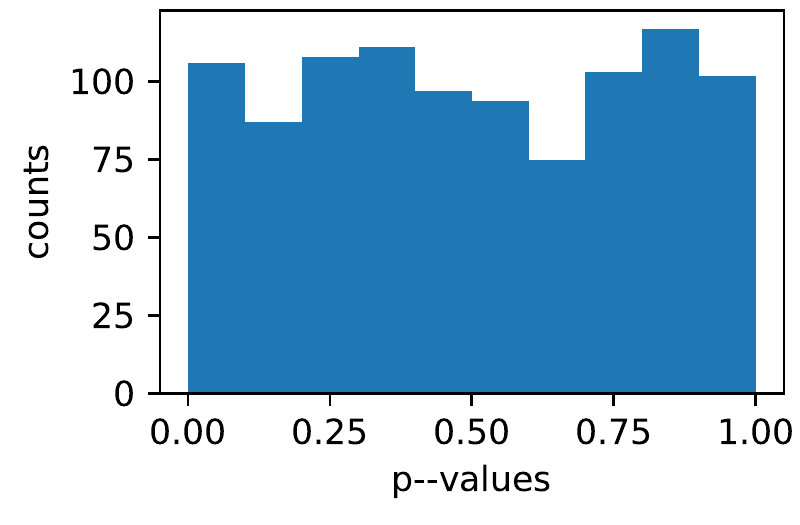}}\\
    (b) \parbox[t][2.1in][c]{3.2in}{%
      \includegraphics{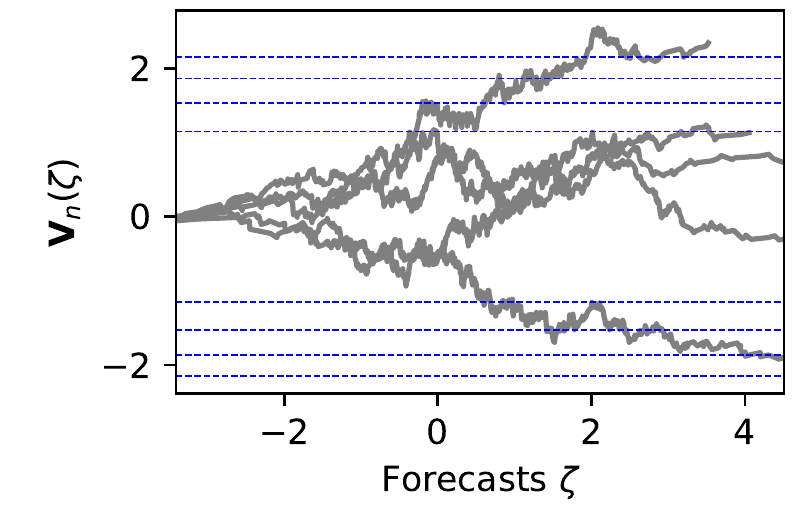}}
  \end{center}
    \caption{\label{fig:quant}
      Test results for quantile forecasts for artificial data.
      The setup is exactly as in Figure~\ref{fig:binary}.
      A Kolmogorov-Smirnov test for uniformity of the histogram gives a $p$-value of~\pquantexp.
            Panel (b)~shows five typical realisations of the process $\epV_n$.
      The dashed lines in Panel~(b) are as in Figure~\ref{fig:binary}.
    }
\end{figure}

The artificial data for our experiments is generated with an autoregressive process of order one.
We define the process $\{X_n, n = 1, 2, \ldots \}$ recursively trough
\beqn{equ:4.10}
X_{n+1} = a X_n + R_{n+1}
\qquad \text{for $n = 0, 1, \ldots$}
\eeq
where $X_0, R_1, R_2, \ldots$ are independent and normally distributed with mean zero and variance $\E(X_0^2) = \frac{1}{1 - a^2}$ and $\E(R^2_k) = 1$ for all $k = 1, 2, \ldots$.
Further, $a = 0.8$ for our experiments.
These choices render the AR-process $\{X_n, n = 1, 2, \ldots \}$ stationary and ergodic.
\paragraph{Binary forecasts}
As binary verification we consider $Y_k = \cf_{\{X_k \geq 0\}} \cdot Z_k + \cf_{\{X_k < 0\}} \cdot (1 - Z_k)$, where $\{Z_k, k = 1, 2, \ldots\}$ are independent and identically distributed Bernoulli variables (also independent from the $\{X_k\}$) with success probability $p_s = 0.95$.
The additional Bernoulli variables $\{Z_k\}$ represent a form of observational noise or confusion of the observables.
As forecasts we use $f_k = \P(Y_k = 1 | X_{k-1})$, which can be calculated explicitly through
\beq{equ:4.20}
f_k = p_s \Gauss(a X_{k-1})
+ (1 - p_s) \* (1 - \Gauss(a X_{k-1})),
\eeq
for $k = 1, 2, \ldots$, where $\Gauss(.)$ denotes the standard normal cumulative distribution function.
The verification-forecast pairs satisfy the reliability hypothesis~\eqref{equ:1.20} because the AR-process is Markov.
Furthermore, the distribution of $X_k$ is continuous (in fact it is normal); it then follows from Equation~\eqref{equ:4.20} that the forecasts $\{f_k, k = 1, 2, \ldots\}$ are stationary and ergodic and have a continuous distribution.
Hence the conditions stated at the beginning of Section~\ref{sec:methodology} are satisfied and hence our test should exhibit the stated asymptotic behaviour.
To verify this, we ran a Monte-Carlo experiment to generate the distribution of the $p$-value.
The Monte-Carlo experiment comprised \nrmcrepeat~runs; in each run, the test was applied to a set of \nrtstamps~verification-forecast pairs (i.e.\ roughly ``two years'' of daily forecasts), then the test statistic and finally the $p$-value was computed.
The runs constituted identical and statistically independent experiments.
The resulting empirical distribution of $p$--values should be uniform, meaning that if used with a significance level of $\theta$, the null hypothesis should be accepted with probability $\theta$.
The results are shown in Figure~\ref{fig:binary}.
Panel (a)~shows a histogram of the $p$-values for each of the \nrmcrepeat~tests.
As discussed, the histogram should exhibit a uniform distribution, up to statistical fluctuations; this is confirmed here through a Kolmogorov-Smirnov test which gives a $p$-value of~\pbinaryexp.
We can thus be confident that the analysis is correct and that a time series of length \nrtstamps~appears sufficient for the limiting distribution to be an acceptable approximation.
Panel~(b) shows five typical realisations of the process $\epV_n$.
The fluctuations of the process are readily apparent, but their magnitude is not homogenous for different values of $\zeta$.
This is because the limiting process is a Wiener process with (nonuniform) re-scaling of time, rather than a pure Wiener process.
The dashed lines in Panel~(b) indicate quantiles for the supremum of $\epV$ for the levels $\frac{1}{2}, \frac{1}{4}, \frac{1}{8}$ and~$\frac{1}{16}$.
That is, the path of $\epV$ exceeds the innermost pair of dashed lines with probability $\frac{1}{2}$, the pair of lines next further out is exceeded with probability $\frac{1}{4}$ and so on, while $\epV$ exceeds the outermost pair with probability~$\frac{1}{16}$.
Plotting these lines gives a direct visual indication as to whether a given path of $\epV$ would be typical under the null hypothesis.
\paragraph{Conditional mean forecasts}
A similar experiment was applied to conditional mean forecasts.
As verification we consider $Y_k = X_k$, that is the current state of the AR-process; forecasts were based on the previous state $X_{k-1}$.
More specifically, we use $f_k := \E(Y_k| X_{k-1}) = a X_{k-1}$.
The verification-forecast pairs satisfy the reliability hypothesis~\eqref{equ:1.20}, again because the AR-process is Markov.
It follows as before that the forecasts $\{f_k, k = 1, 2, \ldots\}$ are stationary and ergodic and have a continuous distribution.
Hence the conditions stated at the beginning of Section~\ref{sec:methodology} are satisfied and hence our test should exhibit the stated asymptotic behaviour.
Again, a Monte-Carlo experiment was run to generate the distribution of the $p$-value, using the same setup as for the binary case.
The results are shown in Figure~\ref{fig:mean}.
Panel~(a) shows the histogram of the $p$--values which up to statistical fluctuations is uniform; a Kolmogorov-Smirnov test which gives a $p$-value of~\pmeanexp.
This confirms that the analysis is correct and that a time series of length \nrtstamps~appears to be sufficient for the limiting distribution to be an acceptable approximation also in this case.
Panel (b)~shows five typical realisations of the process $\epV_n$.
\paragraph{Conditional quantile forecasts}
Finally, a similar experiment was applied to conditional quantile forecasts.
As verification we consider again the current state of the AR-process $Y_k = Z_k$.
As forecasts we used the 70\%-quantile of $Z_k$ conditionally on $Z_{k-1}$.
This can be computed directly using the quantile function of the normal distribution since the noise in our AR-process is normal.
The verification-forecast pairs again satisfy the reliability hypothesis~\eqref{equ:1.20}, and it follows as before that the forecasts $\{f_k, k = 1, 2, \ldots\}$ are stationary and ergodic and have a continuous distribution.
Hence once more the conditions stated at the beginning of Section~\ref{sec:methodology} are satisfied and hence our test should exhibit the stated asymptotic behaviour.
The results of the Monte-Carlo experiment are shown in Figure~\ref{fig:quant}.
Panel~(a) shows the histogram of the $p$--values which up to statistical fluctuations is uniform; a Kolmogorov-Smirnov test which gives a $p$-value of~\pquantexp.
Panel (b)~shows again five typical realisations of the process $\epV_n$.
\subsection{How much data is needed?}
\label{sec:how_much_data}
It needs to be kept in mind that the distribution for the test statistic is only correct asymptotically for long archives of verification--forecast pairs.
So strictly speaking, the asymptotic distribution is never {\em exactly} correct and therefore, the distribution of the $p$--values is never exactly uniform.
As a consequence, increasing the number of Monte-Carlo runs should cause the Kolmogorov-Smirnov test for uniform distribution of the $p$--values to fail at some point, depending on the size of the verification--forecast archive.
We have carried out further experiments for the binary case, the mean forecasts, and the quantile forecasts, varying both the number of Monte-Carlo runs as well as size of the verification--forecast archive.
We will not provide detailed results here but the general message seems to be that the size of the verification--forecast archive is adequate for the binary case as well as the quantile case, but that for the conditional mean forecasts the size of the verification--forecast archive might still be a bit small.
Although the experiment for conditional mean forecasts detailed above do not indicate a deviation of the $p$--value distribution from uniformity, we might have been lucky, since a substantially bigger experiment of 5000~Monte-Carlo runs provides $p$--values with a distribution that appears to deviate from uniformity, according to a Kolmogorov-Smirnov test ($p$--value of~\pMeanExpLargeNorm)
This means that for the current experimental setup at least, a time series of \nrtstamps~data points is potentially too short for the distribution of the test statistic to have sufficiently approached the limiting distribution.
This finding however is strongly dependent on the specific setup.
In particular, it emerges from the proof of the uniform central limit theorem that the decay of correlations in the forecast time series play a role, as do large outliers in the sum~$\epU_n$ (see Eq.~\ref{equ:2.150}).
Outliers cannot happen in the binary case and the quantile case (as the entries in the sum are bounded) but they can in the conditional mean case.
We have therefore repeated the experiment (with 5000~Monte-Carlo runs) for the conditional mean forecasts but with an AR-process featuring bounded noise (uniformly distributed, see next section for details), thus removing the possibility of outliers.
The $p$--values of the test now show no indication of deviation from uniformity, according to a Kolmogorov-Smirnov test ($p$--value of~\pMeanExpLargeUnif), consistent with the above discussion.
%
%
\subsection{Temperature forecasts}
\begin{figure}[p]
  \begin{center}
    %
    \includegraphics[width = 0.48\textwidth]{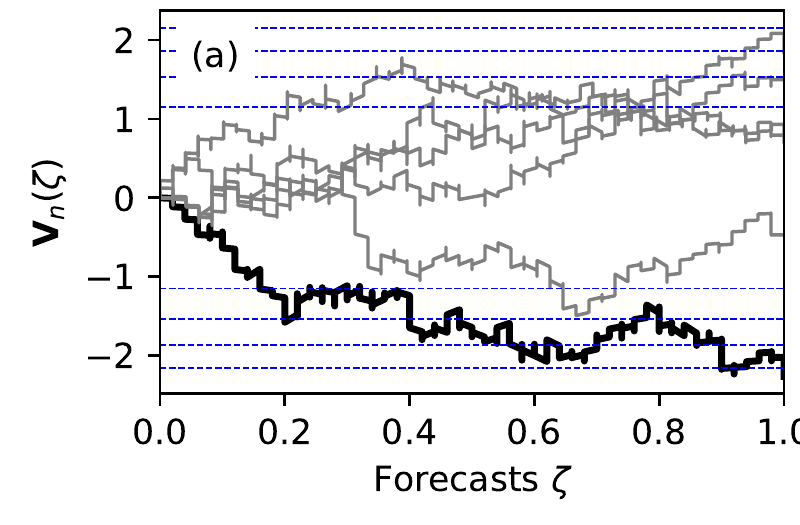}
    \includegraphics[width = 0.48\textwidth]{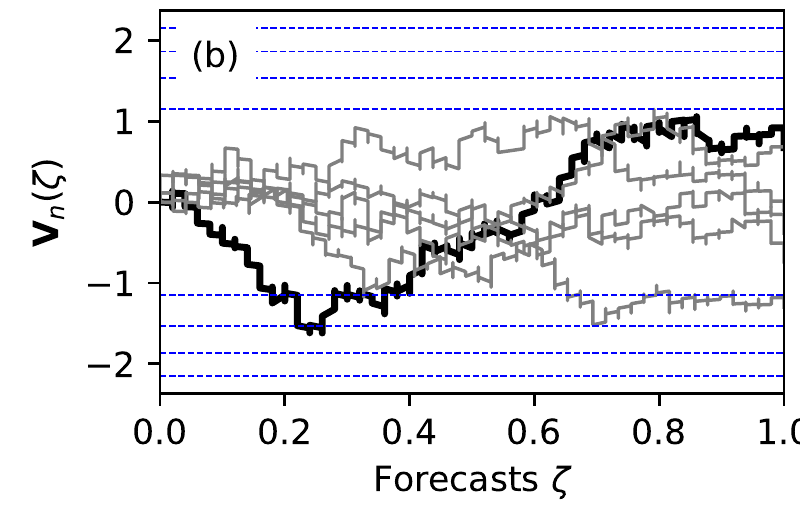}\\
    
    \includegraphics[width = 0.48\textwidth]{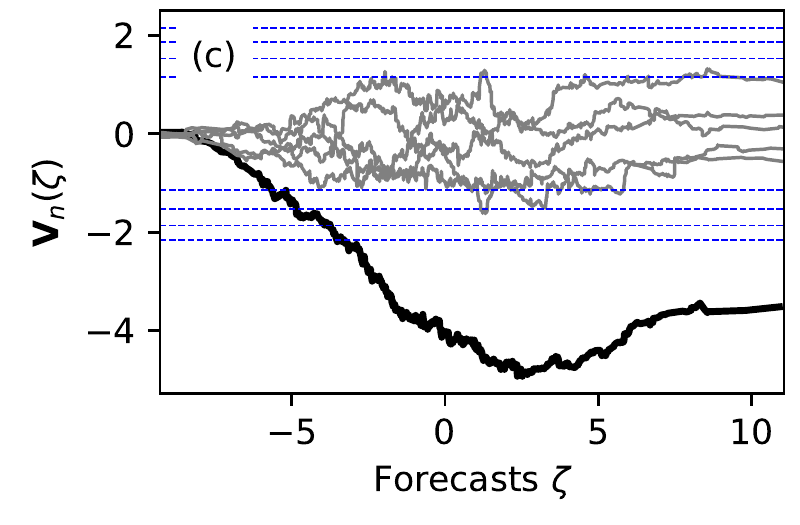}
    \includegraphics[width = 0.48\textwidth]{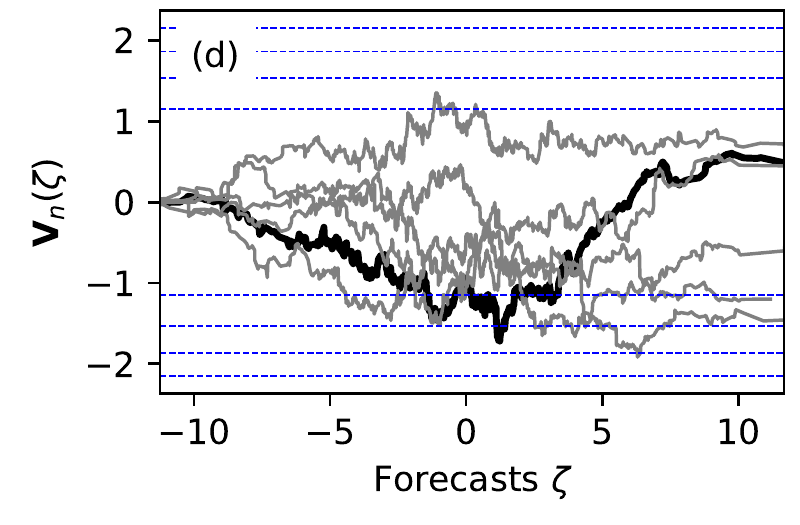}\\
    \includegraphics[width = 0.48\textwidth]{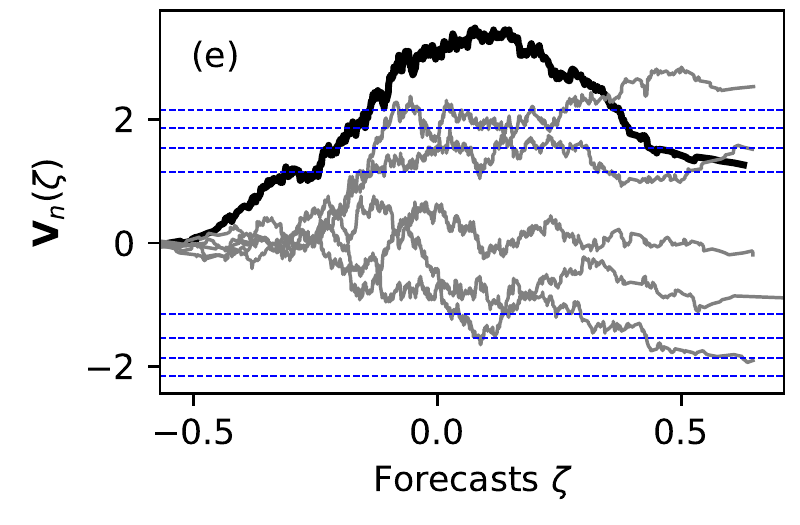}
    \includegraphics[width = 0.48\textwidth]{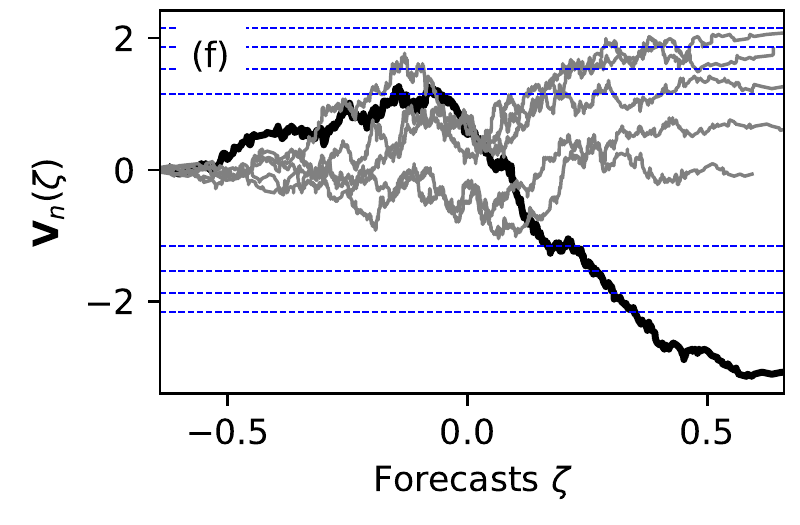}\\
    \end{center}
\caption{\label{fig:paths}
  The realisations of $\epV_n$ (in black) for probability, mean, and quantile forecasts (rows) and two stations (Nuremberg and Bremen; columns), as in Table~\ref{tab:pvals}.
  An additional five realisations of the process $\epV_n$ are shown in grey; these were generated by a bootstrap procedure and show typical behaviour under the reliability hypothesis (see text for details).
The dashed lines show quantiles as in Figure~\ref{fig:binary}.
Probability and mean forecasts for Bremen (panels~(b) and~(d)) show no evidence for deviations from reliability while others do.
Panel (e) (quantile forecast for Nuremberg) shows evidence for deviation from reliability but {\em no} evidence for deviation from unconditional reliability (see text for details).
    }
\end{figure}

%
%
%
\label{sec:ecmwf_experiment}
The proposed methodology was applied to operational weather forecasts.
The verifications comprise temperature measurements from several weather stations in Germany, taken daily at 12UTC.
The forecasts are based on the medium range ensemble prediction system of the European Centre for Medium Range Weather Forecasts (ECMWF)\footnote{We are grateful to ECMWF and Zied Ben Bouall\`{e}gue for kindly providing the data.}.
The system produces ensemble forecasts for the global atmosphere and gets initialised four times a day.
It comprises 50~ensemble members, where each ensemble member represents a possible future evolution of the global atmosphere out to a lead time of 10~days.
For our study however, we will only use the forecasts initialised at 12UTC and with a lead time of 24~hours, or if measured in observation time steps rather than absolute time, the lead time is equal to~1 as per the standing assumption in this paper.
Below we present results for two weather stations
(Bremen and Nuremberg)\footnote{German Weather Service~(DWD) Station~ID's~691 and~3668, respectively}.
We stress that the results are not to be understood as a comprehensive or representative reliability study of the ECMWF ensemble forecasting system, either in its entirety or of parts of it.
The aim of the experiments is merely to demonstrate the feasibility of applying the methodology to operational forecast data, and to confirm that quantitatively the results are plausible.
Apart from a conversion to seasonal anomalies (see below), the forecasts were neither post-processed or re-calibrated in any way.
The verification--forecast pairs cover a period between 1st~of~December 2014 to 30st~of~September 2020 (resulting in about 2130~values).
The verifications and ensembles were converted to anomalies by subtracting a {\em climate normal} of the form
\beqn{equ:4.30}
c(k) = c_1 + c_2 \cos(\omega k) + c_3 \sin(\omega k)
\qquad \text{where}
\qquad \omega = \frac{2 \pi}{365.2425}.
\eeq
The coefficients $c_1, c_2, c_3$ were found by a least squares fit onto the entire set of temperature measurements from the station under concern.
The first~1000 measurements were then used in the actual experiment.
The fact that the climate normal has already been fitted to and subtracted from these values strictly speaking constitutes an in--sample calibration of the data.
This was deemed not to be a problem here though given that the climate normal comprises a low-complexity model and that the experiment is for illustrative purposes only.
The ensemble forecasts were used to generate mean forecasts, (binary) probability forecasts, and quantile forecasts as follows.
We write $\bX_n$ for the entire ensemble at time $n$; this ensemble comprises~50 ensemble members, and the $k$'th ensemble member is written as $X_n^{(k)}$ so that $\bX_n = (X_n^{(1)}, \ldots, X_n^{(K)})$ with $K = 50$ in our case.
The assumption underlying the way we generate mean, probability, and quantile forecasts is basically that the ensemble members at time $n$ are randomly drawn from the forecast distribution, that is the conditional distribution of the verification $Y_n$, given the information available to the forecaster (which, as discussed, includes all forecasts and verifications up to and including time $n-1$).
In particular, the ensemble members are completely exchangeable~\citep[see][for a discussion of this point]{broecker_exchangeability_2011}.
As the ensemble members are real valued in our case, we may assume that they are sorted in ascending order, that is $X_n^{(1)} \leq \ldots \leq X_n^{(K)}$ for each $n = 1, 2, \ldots$ (this will simplify subsequent notation).
We consider probability forecasts for the binary event $Y_n > 0$, that is whether the measured temperature exceeds the climate normal.
The forecast is constructed by a (regularised) frequency estimator, that is we count the relative number of ensemble members exhibiting the same event
\beqn{equ:4.50}
f_n := \frac{N_n + 1/2}{51}
\qquad \text{for~$n = 1, 2, \ldots$,}
\eeq
where $N_n$ is the number of ensemble members $X_n^{(k)}$ such that~$X_n^{(k)} > 0$.
(Our regularisation of $f_n$ amounts to assuming that there is an additional fictitious ensemble member ``split in half'', one half always exhibiting the event while the other one never does.)
It needs to be mentioned that the forecasts only assume a discrete set of 51~values, while our conditions require that the range of forecasts be continuous.
This also causes the paths in Panels~(a) and~(b) in Figure~\ref{fig:paths} to look a bit different from the other panels, as the forecast values do not range continuously over the abscissa.
We have not fully analysed this problem but it seems plausible that it is immaterial, in the sense that the stated mathematical results about the limiting distribution of $\tau_n$ still hold in the limit of an infinitely large ensemble.
Mean forecasts are generated by simply taking the ensemble mean
\beqn{equ:4.40}
f_n := \frac{1}{K} \sum_{k = 1}^K X_n^{(k)}
\qquad \text{for~$n = 1, 2, \ldots$. }
\eeq
Finally, regarding quantile forecasts, we note that the conditional probability of finding the verification~$Y_n$ to be equal to or larger than the~$k$'th ensemble member~$X_n^{(k)}$ is given by $k / (K+1)$~\citep[see for instance][for a proof]{broecker_stratified_serial_dependence_2020}.
We will use $k = 25$ here, that is we take
\beqn{equ:4.55}
f_n := X_n^{(k)}
\qquad \text{with $k=25$ for~$n = 1, 2, \ldots$ }
\eeq
as a quantile forecast with level $\alpha = \frac{25}{51}$.
No attempt was made to re-calibrate these forecasts to improve reliability or to increase the performance in any way, and the following results should be considered with this in mind.
Table~\ref{tab:pvals} contains the $p$--values for the three types of forecasts and the two stations (corresponding to the three rows and the two columns of Table~\ref{tab:pvals}, respectively).
Figure~\ref{fig:paths} shows the realisations of $\epV_n$ in black.
The arrangement of the plot panels is the same as in Table~\ref{tab:pvals}.
From both the Table and the Figure we obtain a mixed picture, with some experiments showing no evidence for deviations from reliability while others do.
Each panel furthermore shows an additional five realisations of the process $\epV_n$.
Each of these realisations was generated by first randomly selecting a set of~1000 forecast instances $\bX_n$ from the entire data set of 2130~instances without replacement, and then substituting, for each $n$, a randomly selected member of the ensemble $\bX_n$ for the verification $Y_n$.
Although a detailed analysis of this approach is left to future work, there is reason to believe that in some appropriate limit, the realisations obtained in this way are distributed as if the forecasts were indeed a reliable forecast for the verification.
This belief is based on two observations.
Firstly, the asymptotic distribution of $\epV_n$ under the null hypothesis would not change if the verification--forecast pairs $\{(Y_n, f_n), n = 1, 2, \ldots\}$ were independent.
(Our argument in Sec.~\ref{sec:methodology} is that as long as the conditions stated there apply, any potential dependence is irrelevant.)
Secondly, in the limit of a very large ensemble, the empirical mean over the ensemble members approaches the conditional mean of any of its members, given the information available at forecast time.
A similar argument can be made for probability and quantile forecasts.
As a consequence, there is reason to believe that the additional five realisations behave roughly like independent draws from the distribution of the process $\epV_n$ {\em under the null hypothesis}.
We stress however that pending a more thorough analysis of this approach, these additional realisations should serve only as a very rough consistency check.
The interpretation of the dashed lines in the panels of Figure~\ref{fig:paths} is the same as in Figures~\ref{fig:binary},~\ref{fig:mean}, and~\ref{fig:quant}.
\begin{table}
  \begin{center}
\begin{tabular}{rll}
 &       Nuremberg & Bremen \\
 Prop.  fc. & $0.0474$ &  $0.2113$ \\
 Mean  fc.  &   $1.6431 \cdot 10^{-6} $ & $0.1714$\\
 Quant. fc. & $0.0010$ & $0.0035$
 %
    \end{tabular}
    \end{center}
  \caption{\label{tab:pvals}%
    The $p$--values for probability, mean, and quantile forecasts (rows) and two stations (Nuremberg and Bremen; columns).
    Probability and mean forecasts for Bremen (2nd column, rows~1,~2) show no evidence for deviations from reliability while others do.
  }
\end{table}
Although the results must not be interpreted as a representative reliability analysis of this forecasting system, they demonstrate the advantage of testing reliability {\em uniformly and simultaneously} over all values of the forecast.
Indeed, for an unconditional test one would consider the sum in Equation~\eqref{equ:2.20} (or Eqs.~\ref{equ:2.80},\ref{equ:2.150} for mean and quantile forecasts, resp.) but with the term $\cf\{f_k \leq \zeta \}$ omitted.
In other words, such a test would merely look at the value of the function $\epV_n$ at the end point $\zeta_{\infty}$ (where $\zeta_{\infty} = 1$ for probability forecasts and $\zeta_{\infty} = \infty$ for mean forecasts and quantile forecasts), rather than at the entire function for all $\zeta$.
As discussed in Section~\ref{sec:methodology}, under the hypothesis of reliability, $\epV_n(\zeta_{\infty})$ has a standard normal distribution, and it turns out that the blue lines in Figure~\ref{fig:paths} delineate regions with probability approximately $0.2506, 0.1250, 0.0625,$ and~$0.0313$ with respect to the standard normal distribution.
Considering quantile forecasts for Bremen for instance (Fig.~\ref{fig:paths}, panel~e), a test for unconditional reliability based on $\epV_n(\zeta_{\infty})$ would provide a $p$-value of at least $0.1250$ (judging from the plot), while the conditional test based on the entire path gives a $p$-value of about $0.001$ (see Tab.~\ref{tab:pvals}).
The supremum of $|\epV_n(\zeta)|$ is assumed around $\zeta = 0$, with the process increasing for $\zeta < 0$ and then decreasing.
Given the way $\epV_n$ is calculated (see Eq.~\ref{equ:2.150}), we see from Fig.~\ref{fig:paths}(e) that the event $\{Y_k < f_k\}$ happens too frequently as long as $f_k < 0$, but too rarely if $f_k > 0$.
These effects however cancel out on average over all forecasts, meaning that the overall frequency of the event $\{Y_k < f_k\}$ is about $1/2$ which it is expected to be under reliability.
Therefore, the lack of reliability in the present example would go undetected by an unconditional test.
%
%
\section{Power considerations}
\label{sec:power}
\begin{figure}[p]
  %
  \begin{center}
    (a) \parbox[t][2.1in][c]{3.2in}{%
      \includegraphics{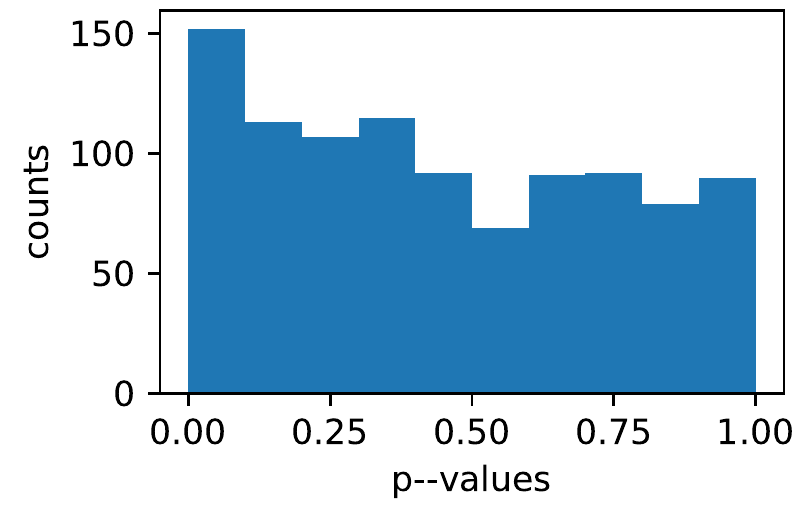}}\\
    (b) \parbox[t][2.1in][c]{3.2in}{%
      \includegraphics{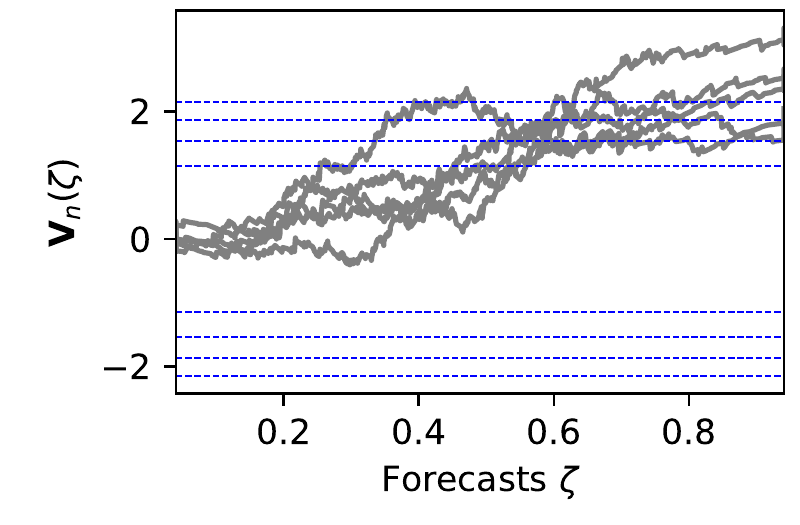}}
  \end{center}
    \caption{\label{fig:bias_binary}
      Panel (a)~shows a histogram of the $p$-values for binary forecasts which are not reliable (deviation from reliability about~3\%, see text).
      This histogram is evidently skewed to smaller values (Kolmogorov-Smirnov test $p$-value is~\pbiasbinaryexp).
      This confirms that a time series of length \nrtstamps~appears to be sufficient for the test to develop appreciable power against this alternative.
      Panel~(b) shows five typical realisations of the process $\epV_n$ (the deviation from reliability is larger than for Panel~(a) for illustrative purposes).
      The dashed lines in Panel~(b) are as in Figure~\ref{fig:binary}.
    }
\end{figure}

\begin{figure}[p]
  %
  \begin{center}
    (a) \parbox[t][2.1in][c]{3.2in}{%
      \includegraphics{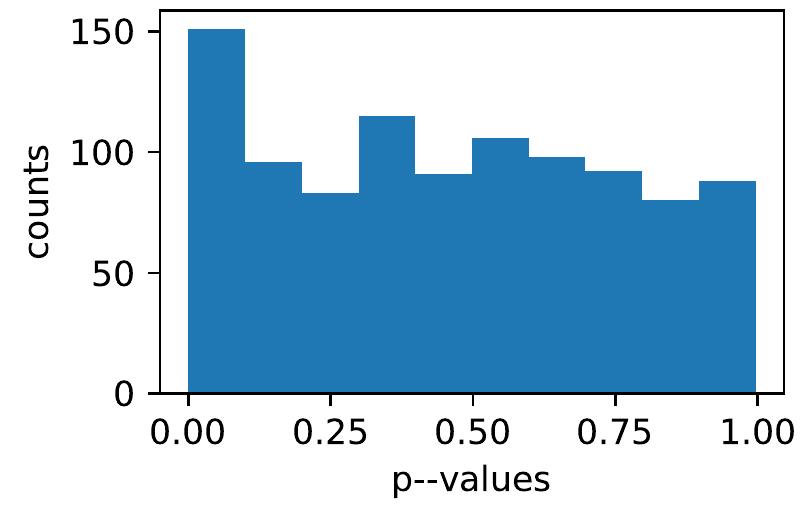}}\\
    (b) \parbox[t][2.1in][c]{3.2in}{%
      \includegraphics{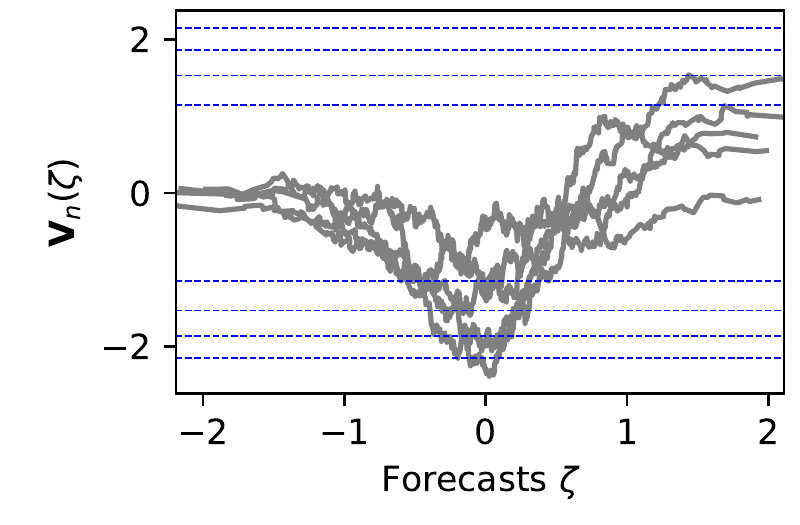}}
  \end{center}
    \caption{\label{fig:bias_mean}
      Panel (a)~shows a histogram of the $p$-values for mean forecasts which are not reliable (deviation from reliability about~2.6\%, see text).
      This histogram is evidently skewed to smaller values (Kolmogorov-Smirnov test $p$-value is~\pbiasmeanexp).
      Hence the test has power against this alternative.
Panel~(b) shows five typical realisations of the process $\epV_n$ (the deviation from reliability is larger than for Panel~(a) for illustrative purposes).
      The dashed lines in Panel~(b) are as in Figure~\ref{fig:binary}.
    }
\end{figure}

\begin{figure}[p]
  %
  \begin{center}
    (a) \parbox[t][2.1in][c]{3.2in}{%
      \includegraphics{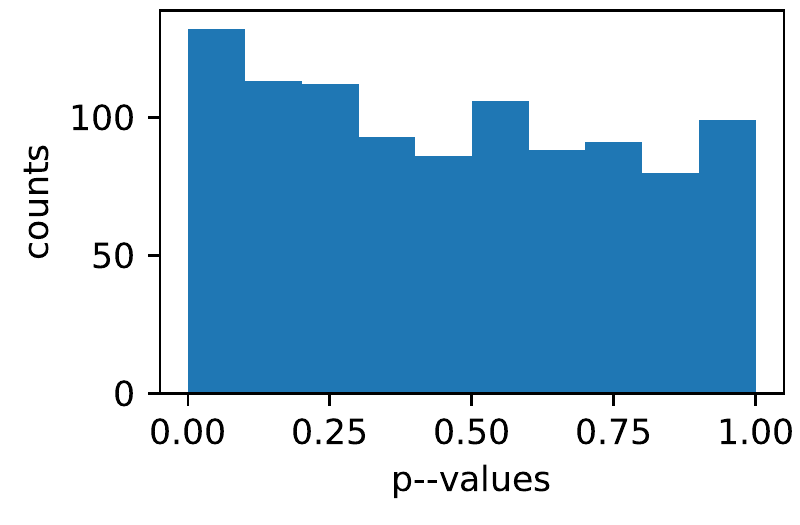}}\\
    (b) \parbox[t][2.1in][c]{3.2in}{%
      \includegraphics{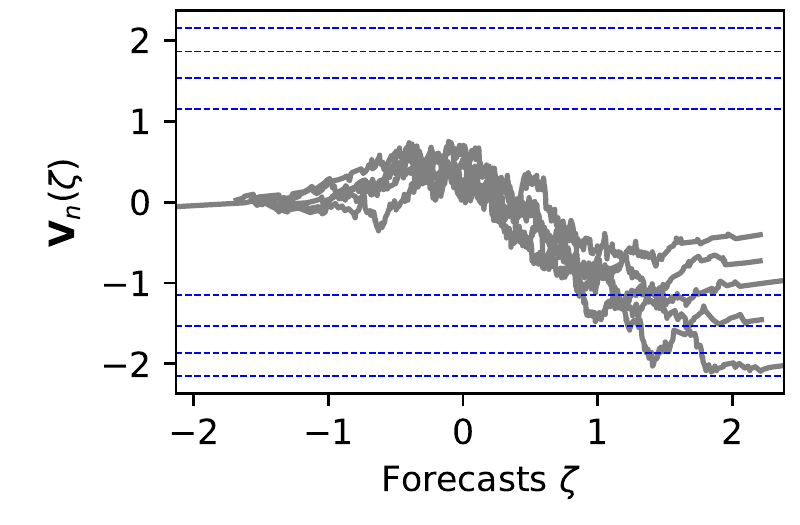}}
  \end{center}
    \caption{\label{fig:bias_quant}
            Panel (a)~shows a histogram of the $p$-values for quantile forecasts which are not reliable (deviation from reliability about~2.4\%, see text).
      This histogram is evidently skewed towards smaller values (Kolmogorov-Smirnov test $p$-value is~\pbiasquantexp).
      This confirms that a time series of length \nrtstamps~appears to be sufficient for the test to develop appreciable power against this alternative.
      Panel~(b) shows five typical realisations of the process $\epV_n$ (the deviation from reliability is larger than for Panel~(a) for illustrative purposes).
      The dashed lines in Panel~(b) are as in Figure~\ref{fig:binary}.
    }
\end{figure}

%
As said previously, the tests are only able to develop power against violations of the relation~\eqref{equ:1.10} (or relations~\eqref{equ:1.43} and~\eqref{equ:1.47} in the case of mean and quantile forecasts, respectively).
We will first demonstrate this from a theoretical perspective and carry out a few experiments afterwards.
Our theoretical analysis will focus on the binary case as the considerations for mean and quantile forecasts are very similar.
We will write $\{g_k, k = 1, 2, \ldots\}$ for a potentially unreliable set of forecasts corresponding to the verifications $\{Y_k, k = 1, 2, \ldots\}$.
We observe that
\beq{equ:5.05}
\P(Y_k = 1 | g_k) = g_k + \psi(g_k)
\qquad \text{for $k = 1, 2, \ldots$}
\eeq
for some function $\psi$, simply because $\P(Y_k = 1 | g_k)$ is always a function of $g_k$.
The hypothesis~\eqref{equ:1.10}, if in force, would imply that $\psi = 0$ identically.
Now we assume this to be no longer the case.
More specifically, we assume that there exists a $\zeta_* \in [0, 1]$ such that\footnote{%
  It is possible to show that Equation~\eqref{equ:5.07} is equivalent to $\psi$ being nonzero with nonzero probability, or again equivalently that there is some $\epsilon > 0$ and a set $A \subset [0, 1] $ which contains forecasts with positive probability such that either $\psi(x) \geq \epsilon$ for all $x \in A$, or $\psi(x) \leq -\epsilon$ for all $x \in A$.}
\beq{equ:5.07}
\E(\psi(g_k) \cf_{\{g_k \leq \zeta_*\}}) = \eta \neq 0.
\eeq
This expectation value does not depend on $k$ if we assume stationarity.
The random function $\epU_n$ now reads as
\beqn{equ:5.09}
\begin{split}
\epU_n(\zeta)
 & = \frac{1}{n} \sum_{k = 1}^n (Y_k - g_k) \* \cf_{\{g_k \leq \zeta\}} \\
& = \frac{1}{n} \sum_{k = 1}^n (Y_k - (g_k + \psi(g_k)) \* \cf_{\{g_k \leq \zeta\}} \\
& \quad + \frac{1}{n} \sum_{k = 1}^n \psi(g_k) \* \cf_{\{g_k \leq \zeta\}}\\
& =: \epU^{(1)}_n(\zeta) + \epU^{(2)}_n(\zeta).
 \end{split}
\eeq
For the first term on the right hand side, we have again a uniform central limit theorem, and in particular, the conclusions in Equations~(\ref{equ:2.50},\ref{equ:2.170}) still hold for $\epU^{(1)}_n$, essentially because $f_k := g_k + \psi(g_k)$ is now a reliable forecast.
For the second contribution however we have 
\beqn{equ:5.010}
\epU^{(2)}_n(\zeta)
\to \E(\psi(g_k) \cf_{\{g_k \leq \zeta\}})
\eeq
by the law of large numbers, and in view of Equation~\eqref{equ:5.07}, we see that $\sqrt{n}\epU^{(2)}_n(\zeta_*)$ behaves like $\sqrt{n} \eta$, which diverges to either $+\infty$ or $-\infty$, depending on the sign of $\eta$.
As a result, the test statistic $\tau_n$ will diverge to $\infty$ and the hypothesis will be rejected with increasing probability as $n \to \infty$.
This demonstrates that the test will exhibit asymptotically unit power against any alternative of the form~\eqref{equ:5.05} (unless $\psi$ is zero with probability one).
We will now discuss a few numerical experiments, confirming that the tests exhibit power against even relatively small deviations from reliability.
We expect the tests to report typically small $p$-values, in the sense that under the alternative hypotheses, the distribution of $p$-values is no longer uniform but skewed to smaller values.
We have seen that failure of the distribution to reach its asymptotic form might also cause the $p$-values to have a non-uniform distribution.
For the experiments considered throughout this paper though, this problem could be remedied by avoiding outliers in the data.
We thus guarantee that the hypotheses are not rejected just because the distribution of the test statistic is still too distinct from its asymptotic limit.
We will therefore use an AR-process with bounded noise in this section, that is, we define the process $\{X_n, n = 1, 2, \ldots \}$ recursively trough
\beqn{equ:5.10}
X_{n+1} = a X_n + R_{n+1}
\qquad \text{for $n = 0, 1, \ldots$}
\eeq
where now $R_1, R_2, \ldots$ are independent and {\em uniformly} (rather than normally) distributed on the interval $[-1, 1]$.
As before, $a = 0.8$ for our experiments, and again the AR-process is asymptotically stationary and ergodic.
\paragraph{Binary forecasts}
The binary verification is defined as before (including observational noise as already described).
As forecasts we use $g_k := \phi_{\epsilon}(f_k)$ for $k = 1, 2, \ldots$, where $f_k$ is the reliable forecast described in Equation~\eqref{equ:4.20} and investigated in the previous Section.
The mapping $\phi_{\epsilon}$ acts as a distortion of the forecast and is given by

\beq{equ:5.15}
\phi_{\epsilon}(x) = x - \epsilon \frac{x}{1 + x^2}
\eeq
for some $\epsilon \geq 0$; note that $\phi_{\epsilon}(x) = x$ for $\epsilon = 0$.
This means that $\epsilon$ can be interpreted as a degree of non-reliability; for $\epsilon > 0$, the forecasts $\{g_k, k = 1, 2, \ldots\}$ are no longer reliable and Equation~\eqref{equ:1.10} is violated for these forecasts.
In fact, we have
\beqn{equ:5.20}
\P(Y_k = 1 | g_k = p) = \phi^{-1}_{\epsilon}(p)
\qquad \text{for all $k = 1, 2, \ldots$},
\eeq
where $\phi^{-1}_{\epsilon}$ is the inverse of $\phi_{\epsilon}$.
(We note that $\phi^{-1}_{\epsilon}(p)$ will be in the unit interval if $p$ is, which is necessary for the $g_k$ to be interpretable as probability forecasts.)
For the numerical experiments, we use the value $\epsilon = 0.05$, that is a very small deviation from reliability.
A more informative measure of deviation from reliability is the relative mean square deviation between $g_k$ and $f_k$, more precisely
\beqn{equ:5.30}
\rho := \frac{\sqrt{\E (g_k - f_k)^2}}{\sqrt{\E (f_k - \E f_k)^2} },
\eeq
which due to stationarity does not depend on $k$.
We estimate this quantity through empirical averages
\beqn{equ:5.40}
\rho_n := \frac{\sqrt{\overline{(g_k - f_k)^2}}}{\sqrt{\overline{(f_k - \overline{f_k})^2}}}.
\eeq
This quantity turns out to have values less than $0.05$, see below.
As in the last section, we carried out a Monte-Carlo experiment to generate the distribution of the $p$-value.
The results are shown in Figure~\ref{fig:bias_binary}.
Panel~(a) shows a histogram of the $p$-values for each of the \nrmcrepeat~tests.
It is already visually apparent that this histogram is not uniform but skewed to smaller values, and indeed a Kolmogorov-Smirnov test gives a $p$-value of~\pbiasbinaryexp.
This confirms that a time series of length~\nrtstamps{} is sufficient in order that the probability of rejection is significantly elevated, or in other words that the test has power against this alternative.
Of course, the effect is still relatively small and in order to see rejection with virtual certainty, longer data sets are needed.
One has to keep in mind though that the deviation from reliability is still very small, with our estimate of $\rho$ being $\rho_n = 0.032$.
Panel~(b) shows five realisations of the process $\epV_n$.
For these realisations, we have chosen a larger $\epsilon$ of $0.2$ for illustrative purposes.
\paragraph{Conditional mean and quantile forecasts}
The experiments for conditional mean and quantile forecasts are very similar, and we are able to draw similar conclusions.
The verification is defined as before.
As forecasts we use $g_k := \phi_{\epsilon}(f_k)$ for $k = 1, 2, \ldots$, where $f_k$ is the reliable conditional mean forecast investigated in the previous Section.
The mapping $\phi_{\epsilon}$ is as defined in Equation~\eqref{equ:5.15}.
We note that $\phi(x) \simeq x$ for large $x$, which means that if $|f_k|$ is large, we have approximately $f_k = g_k$ or in other words forecasts with large magnitude remain reliable.
Thereby, we avoid that the test power against this alternative is due to very few instances with large amplitude forecasts.
For the numerical experiments, we use $\epsilon = 0.05$ which results in a value of $\rho_n = 0.026$ for our relative root-mean-square deviation from reliability.
Figure~\ref{fig:bias_mean} shows the results of the Monte-Carlo experiment.
The histogram in panel~(a) is clearly not uniform but skewed to smaller values (Kolmogorov-Smirnov $p$-value of~\pbiasmeanexp).
Hence the test has power against this alternative.
Panel~(b) shows five typical realisations of the process $\epV_n$.
The setup for conditional quantile forecast is exactly the same.
We use $\epsilon = 0.05$ which results in a relative root-mean-square deviation of $\rho_n = 0.024$ from reliability.
Figure~\ref{fig:bias_quant} shows the results of the Monte-Carlo experiment.
The histogram in panel~(a) is clearly not uniform but skewed towards smaller values (Kolmogorov-Smirnov $p$-value of~\pbiasquantexp), whence we can conclude that the test has power against this alternative.
Panel~(b) shows five typical realisations of the process $\epV_n$.
%
%
%
\section{Conclusions}
\label{sec:conclusions}
Reliability of forecasting systems can be considered as a statistical hypothesis that typically refers not to a finite dimensional parameter but to an entire functional relationship.
Attempts to estimate deviations from calibration at a specific forecast value will thus be difficult if the probability of the forecast assuming that value is zero, a difficulty that has been long noted.
In this paper, a rigorous testing methodology was presented that is able to cope with this difficulty, provided the forecasts always verify at the next time step in the future, or are of unit lead time.
The asymptotic distribution of the test statistics was discussed and turns out to be universal; furthermore, the tests develop power against a wide class of alternatives.
Numerical experiments for both artificial data as well as operational weather forecasting systems confirm the feasibility of the methodology.
Regarding potential avenues of future research, we will briefly discuss forecasts for larger lead times, and assessment of multi--dimensional forecasts.
Regarding forecasts for larger lead times, it is clear that the restriction to unit lead time is a severe one, and an extension to larger lead times is needed.
The main difficulties have been mentioned at the end of Section~\ref{sec:reliability} already.
These difficulties are not insurmountable though.
The uniform central limit theorem (Theorem~\ref{thm:a.10} in Appendix~\ref{apx:uclt}) holds in case of larger lead times as well, provided several technical assumptions are imposed.
(The proof will be presented elsewhere as it requires a number of modifications and is substantially longer.)
Unfortunately, the limiting process, although Gaussian, will not have the simple representation in terms of a Wiener process as we have encounter here for unit lead time.
The correlation structure of that process will be more complicated and depend on the correlation structure of the time series of verification--forecast pairs.
As already discussed at the end of Section~\ref{sec:reliability}, the reliability hypothesis provides a lot more information on that correlation structure in the case of lead time $L=1$ than for larger lead times.
This means that for larger lead times, more information regarding this correlation structure will have to be estimated from the data itself and factored into the test statistics, most likely requiring further assumptions.
Regarding higher dimensional forecasts, for instance conditional mean forecasts for multi--dimensional verifications, again more work is needed to extend the presented methodology to that situation.
The difficulties are broadly similar to those one would encounter for higher lead times, although they seem to be easier to resolve in the case of higher dimensional forecasts but for unit lead time.
The uniform central limit theorem (Theorem~\ref{thm:a.10} in Appendix~\ref{apx:uclt}) also holds in case of higher dimensional forecasts, although again the proof will require some modifications.
The limiting process has a relatively simple correlation structure and we conjecture that it can be represented in terms of a multi--parameter Wiener process or Brownian sheet.
But again some information regarding this correlation structure will have to be estimated from the data itself, although probably not as much as in the case of larger lead times.
\subsection*{Acknowledgement}
Fruitful discussions with Cl\'{e}ment Dombry, Leonard~A.~Smith and Tobias Kuna are gratefully acknowledged.
  Forecast and verification data for an example were kindfully provided by the European Centre for Medium Range Weather Forecasting. 
  \appendix
%
%
\section{Uniform law of large numbers and central limit theorems}
\label{apx:uclt}
The aim of this section is to give a precise statement of what was called the uniform central limit theorem throughout the paper.
The statement will be general enough so as to cover the cases of probability forecasts, mean forecasts, and quantile forecasts, but it likely applies to other types of forecasts as well.
A proof will also be provided.
Intended for the interested reader, it is kept rather short.
Let $(\Omega, \cF, \P)$ be a probability space and $\{(\phi_k, f_k), k \in \Z\}$ be a stochastic process, where both components $\phi_k$ and $f_k$ are real numbers for all $k \in \Z$.
Although the tests depend on $(\phi_n, f_n)$ only for positive $n$, assuming that these data form a bi--infinite stationary and ergodic process will simplify the subsequent discussion.
In the cases of probability forecasts and mean forecasts we have $\phi_k = Y_k - f_k$, while for quantile forecasts (of some level $\alpha$) we have $\phi_k = \cf_{\{Y_k < f_k\}} - \alpha$.
Our aim is to prove that under the hypothesis of reliability and additional technical conditions, the random function $\epU_n$, where
\beqn{equ:a.20}
\epU(\zeta) := \frac{1}{\sqrt{n}} \sum_{k = 1}^n \phi_k \cf_{\{f_k \leq \zeta\}},
\eeq
converges in distribution to Brownian motion with re-scaled time (in a suitable space of continuous functions).
To formalise the statement and the conditions, we define the filtration $\{\cF_n, n \in \Z \}$ through $\cF_n := \sigma\{(\phi_k, f_{k+1}), k \leq n \}$.
With respect to these objects, we impose
\begin{assumption}
  \label{thm:a.10}
  \begin{enumerate}
  \item \label{thm:a.10.1}
  The process $\{(\phi_k, f_k), k \in \Z\}$ is stationary and ergodic.
  \item \label{thm:a.10.2}
We have
\beq{equ:a.10}
\E(\phi_n | \cF_{n-1}) = 0
\qquad \text{for $n \in \Z$.}
\eeq
  \item \label{thm:a.10.3}
    There is a sigma~finite measure $\lambda$ on $\R$ without atoms so that the conditional distribution of $f_1$ given $\cF_{-1}$ has a density with respect to $\lambda$, that is, there exists a measurable function
    \beqn{equ:a.30}
    p:(\R, \cB_{\R}) \times (\Omega, \cF_{-1}) \to \R_{\geq 0}
    \eeq
    so that $(B, \omega) \to \int_B p(x, \omega) \lambda(\dd x)$ is a regular version of $\P(f_1 \in . | \cF_{-1})$.
  \item \label{thm:a.10.4}
    For some $r > 1$ we have the integrability conditions
    \beqn{equ:a.35}
    \E(|\phi|^{4 r} \cdot p(f_1, \omega) ) ) < \infty
    \eeq
    and
    \beqn{equ:a.37}
    \E(|\phi|^{4 r} ) < \infty
    \eeq
\end{enumerate}
  %
  %
\end{assumption}
Of course, Equation~\eqref{equ:a.10} is the reliability assumption (Eqs.~\ref{equ:1.20},\ref{equ:1.30},\ref{equ:1.40} for the different types of forecasts).
Finally, define the function $G:\R \to \R, \zeta \to \E(\phi^2_1 \cf\{f_1 \leq \zeta\})$.
\begin{theorem}
  \label{thm:a.20}
  If Assumption~\ref{thm:a.10} holds, then $\epU_n$ converges in distribution to $W \circ G$, the Brownian motion composed with $G$, in the space $D(\R)$ of c\`{a}dl\`{a}g functions with the Skorokhod topology.
  Furthermore $W \circ G$ is continuous, and therefore the convergence is in distribution with respect to the topology of uniform convergence, too.
\end{theorem}
The proof will occupy the remainder of this section.
For the first part, according to~\citet{billingsley_convergence_probability_measures_1999}, Theorem~13.5, we have to prove that
for any finite collection $-\infty < t_1 < \ldots < t_k < \infty$ we have a central limit theorem for the finite dimensional distributions, that is
  \beq{equ:a.40}
  (\epU_n(t_1), \ldots, \epU_n(t_k)) \to (W \circ G(t_1), \ldots, W \circ G (t_k));
  \eeq
and further that there exists a non-decreasing, continuous, and bounded function $L: \R \to \R$ and some $\varkappa > 0$ such that for any $-\infty < t_1 \leq t_2 \leq t_3 < \infty$ and $n \in \N$ we have
  \beq{equ:a.50}
  \E \left[
    (\epU_n(t_2) - \epU_n(t_1))^2
    (\epU_n(t_3) - \epU_n(t_2))^2
    \right]
    \leq (L(t_3) - L(t_1))^{1 + \varkappa}.
  \eeq
For the second part, we need to show that $G$ is continuous.
This implies that $W \circ G$ is continuous with probability one.
Since the Skorokhod topology relativised to the set of continuous functions agrees with the uniform topology, the convergence takes place in distribution with respect to the uniform topology as well.
Regarding the central limit theorem for the finite dimensional distributions (Eq.~\ref{equ:a.40}), it follows from Assumption~\ref{thm:a.10}.\ref{thm:a.10.1} and~\ref{thm:a.10}.\ref{thm:a.10.2} that 
for any finite collection $-\infty < t_1 < \ldots < t_k < \infty$ the process
$\{(\sqrt{n} \epU_n(t_1), \ldots, \sqrt{n} \epU_n(t_k)), n \in \N\}$ is a martingale with stationary and ergodic increments.
A direct calculation will show that the covariance is independent of $n$ and agrees with that of $(W \circ G(t_1), \ldots, W \circ G (t_k))$.
The statement now follows from standard Martingale central limit theorems, such as~\citet{vaart_time_series_2010}, Theorem~4.17 and the Cram\'{e}r--Wold device.
To prove Equation~\eqref{equ:a.50}, we follow a strategy similar to that of~\citet{koul_nonparametric_model_checks_1999}, Lemma~3.1.
For $t_1 \leq t_2 \leq t_3$ fixed and $j \in \N$ we define $A_j := \phi_j \cf_{\{t_1 < f_j \leq t_2\}}$ and $B_j := \phi_j \cf_{\{t_2 < f_j \leq t_3\}}$.
With these definitions, we may now write
  \beq{equ:a.60}
  \E \left[
    (\epU_n(t_2) - \epU_n(t_1))^2
    (\epU_n(t_3) - \epU_n(t_2))^2
    \right]
  = \sum_{i, j, k, l = 1}^n \E(A_i A_j B_k B_l), 
  \eeq
  Since $A_j B_j = 0$ for all $j \in \N$, we have in particular that the terms in the sum with $\max \{ i, j\} = \max \{ k, l\}$ vanish.
  We will thus focus on terms with $\max \{ i, j\} > \max \{ k, l\}$; terms where $\max \{ i, j\} < \max \{ k, l\}$ can be treated in the same way.
  If in addition $i > j$, then $i > \max \{ j, k, l \}$, and $\E(A_i A_j B_k B_l) = 0$ due to Assumption~\ref{thm:a.10}.\ref{thm:a.10.2}.
  We can thus focus on terms of the form $\E(A^2_i B_k B_l)$ with $i > \max \{ k, l\}$.
  Keeping $i$ fixed for now but summing over $k, l$ and using a simple estimate, we arrive at
  \beq{equ:a.70}
  \begin{split}
  \sum_{k, l = 1}^{i-1} \E(A^2_i B_k B_l)
   & = \E (A^2_i \cdot (\sum_{k = 1}^{i-1} B_k)^2 )  \\
   & \leq 2\E (A^2_i \cdot B_{i-1}^2 ) + 2 \E (A^2_i \cdot (\sum_{k = 1}^{i-2} B_k)^2 ) 
  \end{split}
  \eeq
  We will deal with the second term first.
  Through a monotone class argument, it can be shown that for every $0 \leq \rho \leq t$, there exists a measurable function $\Phi_{\rho}:(\R, \cB_{\R}) \times (\Omega, \cF_{-1}) \to \R_{\geq 0}$ so that
  \beq{equ:a.80}
  \E(|\phi_1|^{\rho} | f_{1}, \cF_{-1}) = \Phi_{\rho}(f_{1}, \omega)
  \eeq
  almost surely.
  By stationarity, the second term in Equation~\eqref{equ:a.70} is equal to $\E (A^2_1 \cdot (\sum_{k = 2 - i}^{-1} B_k)^2 )$ and using the definition of $A_1$, the representation~\eqref{equ:a.80} and Assumption~\ref{thm:a.10}.\ref{thm:a.10.3} we have
  \beq{equ:a.90}
  \begin{split}
  \E (A^2_1 \cdot (\sum_{k = 2 - i}^{-1} B_k)^2 )
  & = \E (\phi_1^2
  \cdot \cf \{t_1 < f_{1} \leq t_2\}
  \cdot (\sum_{k = 2 - i}^{-1} B_k)^2 ) \\
  & = \E ( \E ( \phi_1^2 |  f_{1}, \cF_{-1} )
  \cdot \cf \{t_1 < f_{1} \leq t_2\}
  \cdot (\sum_{k = 2 - i}^{-1} B_k)^2 ) \\
  & \leq \E ( \Phi_{2}(f_{1}, \omega)
  \cdot \cf \{t_1 < f_{1} \leq t_2\}
  \cdot (\sum_{k = 2 - i}^{-1} B_k)^2 ) \\
  & = \E ( \int_{t_1}^{t_2} \! \Phi_{2}(x, \omega)
  p(x, \omega) \, \lambda(\dd x)
  \cdot (\sum_{k = 2 - i}^{-1} B_k)^2 ) \\
  \text{(Fubini's theorem)}
  & = \int_{t_1}^{t_2} \! \E (\Phi_{2}(x, \omega)
  \cdot p(x, \omega) \cdot (\sum_{k = 2 - i}^{-1} B_k)^2 )
  \, \lambda(\dd x)\\
  \text{(C-S inequ.)}
  & \leq \int_{t_1}^{t_2} \!
  \sqrt{\E (\Phi_4(x, \omega) \cdot p(x, \omega)^2)}
  \, \lambda(\dd x) \cdot
  \sqrt{\E((\sum_{k = 2 - i}^{-1} B_k)^4 )} \\
  & \leq  (L_1(t_2) - L_1(t_1)) \cdot
  \sqrt{\E((\sum_{k = 2 - i}^{-1} B_k)^4 )}, 
  \end{split}
  \eeq
  where we have introduced the function
  \[
  L_1: t \to  \int_{-\infty}^{t} \!
  \sqrt{\E (\Phi_4(x, \omega) \cdot p(x, \omega)^2)}
  \, \lambda(\dd x).
  \]
  It follows from Assumptions~\ref{thm:a.10}.\ref{thm:a.10.3} and~\ref{thm:a.10}.\ref{thm:a.10.4} that $L_1$ is continuous, non-decreasing, and bounded.
  To the second factor on the right hand side of Equation~\eqref{equ:a.90} we apply Burkholder's inequality:
  \beq{equ:a.100}
  \sqrt{\E ( (\sum_{k = 2-i}^{-1} B_k)^4 )}
  \leq \sqrt{\E ( (\sum_{k = 2-i}^{-1} B_k^2 ) ^2 )}
  \leq (i-2) \sqrt{\E ( B_1^4 ) }
  \eeq
  Assumption~\ref{thm:a.10}.\ref{thm:a.10.3} can be used again to give
  \beq{equ:a.110}
  \begin{split}
  \E ( B_1^4 )
  & = \E ( \int_{t_2}^{t_3} \Phi_4(x, \omega) p(x, \omega) \lambda(\dd x))\\
  & = \int_{t_2}^{t_3} \E (\Phi_4(x, \omega) p(x, \omega))  \lambda(\dd x) \\
  & = L_2(t_3) - L_2(t_2)
  \end{split}
  \eeq
  where we have introduced the function
  \[
  L_2: t \to  \int_{-\infty}^{t} \!
  \E (\Phi_4(x, \omega) \cdot p(x, \omega))  \, \lambda(\dd x).
  \]
  It follows from Assumptions~\ref{thm:a.10}.\ref{thm:a.10.3} and~\ref{thm:a.10}.\ref{thm:a.10.4} that $L_2$ is continuous, non-decreasing, and bounded.
  In summary, the estimates~(\ref{equ:a.90},\ref{equ:a.100},\ref{equ:a.110}) give
  \beq{equ:a.120}
  \E (A^2_i \cdot (\sum_{k = 1}^{i-2} B_k)^2 )
  \leq C (i - 2) (L_1(t_2) - L_1(t_1)) (L_2(t_3) - L_2(t_2))^{1/2} 
  \eeq
  for the second term in Equation~\eqref{equ:a.70}.
  For the first term in Equation~\eqref{equ:a.70} we use the Cauchy--Schwartz inequality and again stationarity to obtain
  \beq{equ:a.130}
  %
    %
  \E (A^2_1 \cdot B_{0}^2 )
  \leq
  \sqrt{\E \left(
  \phi^{4}_1 \cf\{t_1 < f_1 \leq t_2\} \cf\{t_2 < f_0 \leq t_3\}
  \right)} \cdot \sqrt{ \E (B_0^4) }.
    %
  %
  \eeq
  For the second term on the right hand side, we use Equation~\eqref{equ:a.110}.
  For the first term, we find using H\"{o}lder's inequality with $r$ as in Assumption~\ref{thm:a.10.4} and $s$ so that $\frac{1}{r} + \frac{1}{s} = 1$
  \beq{equ:a.140}
  \begin{split}
  &  \E \left(
  \phi^{4}_1 \cf\{t_1 < f_1 \leq t_2\} \cf\{t_2 < f_0 \leq t_3\}
  \right) \\
  & = \int_{t_1}^{t_2}
  \E (\Phi_{4}(x, \omega ) p(x, \omega ) \cf\{t_2 < f_0 \leq t_3\}) \, \lambda(\dd x) \\
  & \leq \int_{t_1}^{t_2}
  \E (\Phi_{4 r}(x, \omega ) p^{r}(x, \omega ))^{1/r} \, \lambda(\dd x)
  \cdot   \E (\cf\{t_2 < f_0 \leq t_3\})^{1/s}  \\
  & \leq \int_{t_1}^{t_2}
  \E (\Phi_{4 r}(x, \omega ) p^{r}(x, \omega ))^{1/r} \, \lambda(\dd x)
  \cdot \left( \int_{t_2}^{t_3} \E (p(x, \omega)) \lambda(\dd x)  \right)^{1/s}  \\
  & = (L_3(t_2) - L_3(t_1)) \cdot ( L_4(t_3) - L_4(t_2))^{1/s} ,
  \end{split}
  \eeq
  where we have introduced
  $L_3(t) := \int_{-\infty}^{t} \E (\Phi_{4r}(x, \omega ) p^{r}(x, \omega ))^{1/r} \lambda(\dd x)$
  and $L_4(t) := \int_{-\infty}^{t} \E (p(x, \omega)) \lambda(\dd x)$.
  It follows from Assumptions~\ref{thm:a.10}.\ref{thm:a.10.3} and~\ref{thm:a.10}.\ref{thm:a.10.4} that $L_3$ is continuous, non-decreasing, and bounded.
  The same is true for $L_4$ (which is of course just the cumulative distribution function of $f_1$), merely by Assumption~\ref{thm:a.10}.\ref{thm:a.10.3}.
  Combining Equations~(\ref{equ:a.130},\ref{equ:a.140}) we find
  \beq{equ:a.150}
  \begin{split}
  \E (A^2_1 \cdot B_{0}^2 )
  & =  (L_3(t_2) - L_3(t_1))^{1/2}\\
  & \quad \cdot ( L_4(t_3) - L_4(t_2) )^{1/2s} ( L_2(t_3) - L_2(t_2))^{1/2}
  \end{split}
  \eeq
  Combining Equations~\eqref{equ:a.120} and~\eqref{equ:a.150} establishes the condition~\eqref{equ:a.60}.
  It remains to prove the continuity of $G$, but this is evident given the representation
  \beq{equ:a.160}
  G(\zeta) = \int_{-\infty}^{\zeta} \E(\Phi_2(x, \omega) p(x, \omega)) \lambda(\dd x)
  \eeq
  and again Assumptions~\ref{thm:a.10}.\ref{thm:a.10.3} and~\ref{thm:a.10}.\ref{thm:a.10.4}.
  This finishes the proof of Theorem~\ref{thm:a.20}.
%

  %
\end{document}